 \documentclass[final,3p,times]{elsarticle}

\usepackage{amssymb}

\journal{NIMA - Proceedings of the 2024 Advanced Accelerator Concepts workshop}

\begin{document}
\begin{frontmatter}



\title{Effect of Dielectric Wakefields in a Capillary Discharge for Plasma Wakefield Acceleration}


\author[inst1]{L.~Verra}
\author[inst2,inst3,inst4]{M.~Galletti}
\author[inst1]{R.~Pompili}
\author[inst1]{A.~Biagioni}
\author[inst1]{M.~Carillo}
\author[inst2,inst3,inst4]{A.~Cianchi}
\author[inst1]{L.~Crincoli}
\author[inst1,inst5]{A.~Curcio}
\author[inst2]{F.~Demurtas}
\author[inst1]{{G.~{Di~Pirro}}}
\author[inst1]{V.~Lollo}
\author[inst2]{G.~Parise}
\author[inst1]{D.~Pellegrini}
\author[inst1]{S.~Romeo}
\author[inst5]{G.J.~Silvi}
\author[inst1]{F.~Villa}
\author[inst1]{M.~Ferrario}

\affiliation[inst1]{organization={INFN/Laboratori Nazionali di Frascati},
            addressline={Via Enrico Fermi 54}, 
            postcode={00044}, 
            city={Frascati},
            country={Italy}}

\affiliation[inst2]{organization={University of Rome Tor Vergata},
            addressline={Via Della Ricerca Scientifica 1}, 
            postcode={00133}, 
            city={Rome},
            country={Italy}}

\affiliation[inst3]{organization={INFN Tor Vergata},
            addressline={Via Della Ricerca Scientifica 1}, 
            postcode={00133}, 
            city={Rome},
            country={Italy}}

\affiliation[inst4]{organization={NAST Center},
            addressline={Via Della Ricerca Scientifica 1}, 
            postcode={00133}, 
            city={Rome},
            country={Italy}}

\affiliation[inst5]{organization={University of Rome Sapienza},
            addressline={Piazzale Aldo Moro 5}, 
            postcode={00185}, 
            city={Rome},
            country={Italy}}

\begin{abstract}
Dielectric capillaries are widely used to generate plasmas for plasma wakefield acceleration. 
When a relativistic drive bunch travels through a capillary with misaligned trajectory with respect to the capillary axis, it is deflected by the effect of the dielectric transverse wakefields it drives. 
We experimentally show that the deflection effect increases along the bunch and with larger misalignment, and we investigate the decay of dielectric wakefields by measuring the effect on the front of a trailing bunch.
We discuss the implications for the design of a plasma wakefield accelerator based on dielectric capillaries.
\end{abstract}




\end{frontmatter}


\section{Introduction}
Relativistic charged particle bunches generate almost purely transverse space-charge electric field, that can be exploited in high-gradient particle accelerators, such as plasma and dielectric wakefield accelerators (PWFA and DWFA, respectively).

\par In a PWFA, a driver bunch propagates in a low-temperature plasma whose free electrons are set in motion by the space-charge field of the bunch. 
The restoring force provided by the much more massive (and therefore almost immobile) plasma ions generates a periodic oscillation of the plasma electron density, sustaining longitudinal and transverse oscillating fields: the wakefields~\cite{CHEN:1985}.

\par In a DWFA, the space-charge electric field lines terminate on the dielectric surface where they polarize molecules of the material, generating surface charges. 
These surface charges travel synchronously with the bunch but, due to the finite resistivity of the material, they lag slightly behind it~\cite{BANE:1984}. 
Thus, wakefields are generated behind the driver bunch~\cite{BANE:1984,GAI:1988,KEINIGS:1989,NG:1990}.
These wakefields can also be seen as Cherenkov wakefields, since they are generated by the interaction of the space-charge field with a slow-wave structure or medium with a large index of refraction~\cite{BATURIN:2014}.

\par Acceleration of trailing particles or witness bunches with gradients $>$GeV/m was demonstrated with both schemes~\cite{BLUMENFELD:2007,OSHEA:2016}, paving the way for future, compact, high-gradient accelerators. 
In the PWFA case, the generation of high-quality bunches suitable for applications such as free-electron lasers was also demonstrated~\cite{POMPILI:2022,GALLETTI:2022a}.

\par While, in PWFA, the transverse wakefields are generally axisymmetric with respect to the beam propagation axis, i.e., only focusing and defocusing~\cite{KENIGS:1987}, in DWFA they are axisymmetric with respect to the center of the dielectric structure. 
This means that they induce a dipolar effect on the bunch, when the trajectory is not aligned to the structure longitudinal axis~\cite{BATURIN:2014,PARK:2000,BETTONI:2016,CRAIEVICH:2017,OSHEA:2020}. 

The amplitude of the transverse wakefields $W_{\perp}$ along the bunch follows the same trend as the running integral of the bunch charge. 
Thus, particles in the back of the bunch are deflected more strongly than those in the front. The polarity of $W_{\perp}$ is such that the trailing particles are pulled further towards the dielectric material~\cite{CHEN:BOOK}. 
The use of this effect in a “passive streaker” device for time-resolved measurements of the bunch density distribution was proposed in~\cite{BETTONI:2016}.

\par In a recent publication~\cite{VERRA:2024}, we experimentally verified the steering effect in an uncoated dielectric capillary and we showed that the dielectric wakefields can be suppressed by plasma screening of the space-charge field of the bunch.
Here, we report on measurements of the decay time of the dielectric wakefields, obtained by measuring the steering effect on the front of a trailing $witness$ bunch.
We show that, for our experimental conditions, the effect of dielectric wakefields is negligible after $\sim14\,$ps behind the front of the drive bunch. 

\begin{figure}[!ht]
\centering
\includegraphics[width=\columnwidth]{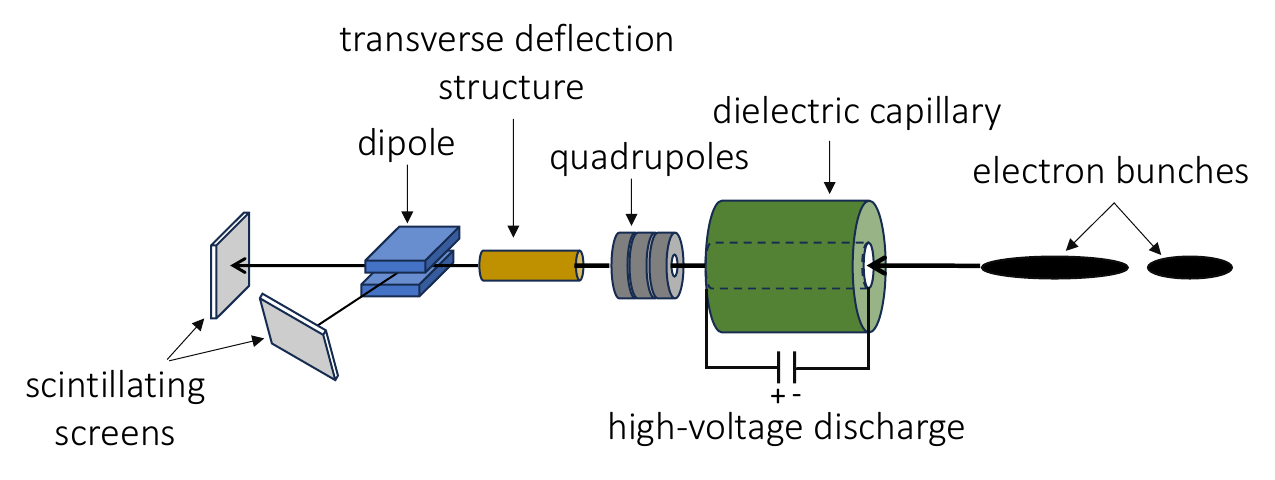}
\caption{Schematic of the SPARC\_LAB experimental setup (not to scale). 
The $e^-$ bunches propagate from right to left.}
\label{fig:1}
\end{figure}

\section{Experimental Setup}

We performed the measurements at the SPARC$\_$LAB facility~\cite{FERRARIO:2013,FERRARIO:2011} (see Fig.~\ref{fig:1}). 
Electron ($e^-$) bunches are generated by illuminating a copper cathode within a radiofrequency (RF) gun with ultraviolet (UV) laser pulses.
Each UV pulse is generated by splitting and delaying the main pulse.
The bunches are accelerated in an RF linac to $E\sim80\,$MeV and focused with transverse root mean square (rms) radius~$\sim 0.4\,$mm at the entrance of a plastic, uncoated capillary with radius $R_c=1\,$mm and length $L=10\,$cm. 
A set of electromagnetic quadrupoles captures the bunches after the interaction and focuses them on a scintillating screen $d=5.3\,$m downstream of the capillary exit for imaging.
A transverse deflection structure (TDS) introduces a head-to-tail vertical correlation to the bunch~\cite{ALESINI:2006}, allowing to obtain time-resolved images of the bunches at the screen position. 
A magnetic dipole and an additional scintillating screen is used to measure the energy of the particles.

\par In the following, we maintain the beam trajectory aligned through the center of the quadrupole magnets, and we record images on the screen with the TDS turned on, and the dipole turned off. 
To study the effect of dielectric wakefields, we shift the capillary with a stepper motor in the horizontal direction (perpendicular to the streaking direction of the TDS) and we measure the induced deflection along the bunches on time-resolved images.
To study the decay time of the dielectric wakefields, we measure the steering effect on the front of the witness bunch while varying its delay with respect to the drive bunch. 
\begin{figure}[ht!]
\centering
\includegraphics[width=\columnwidth]{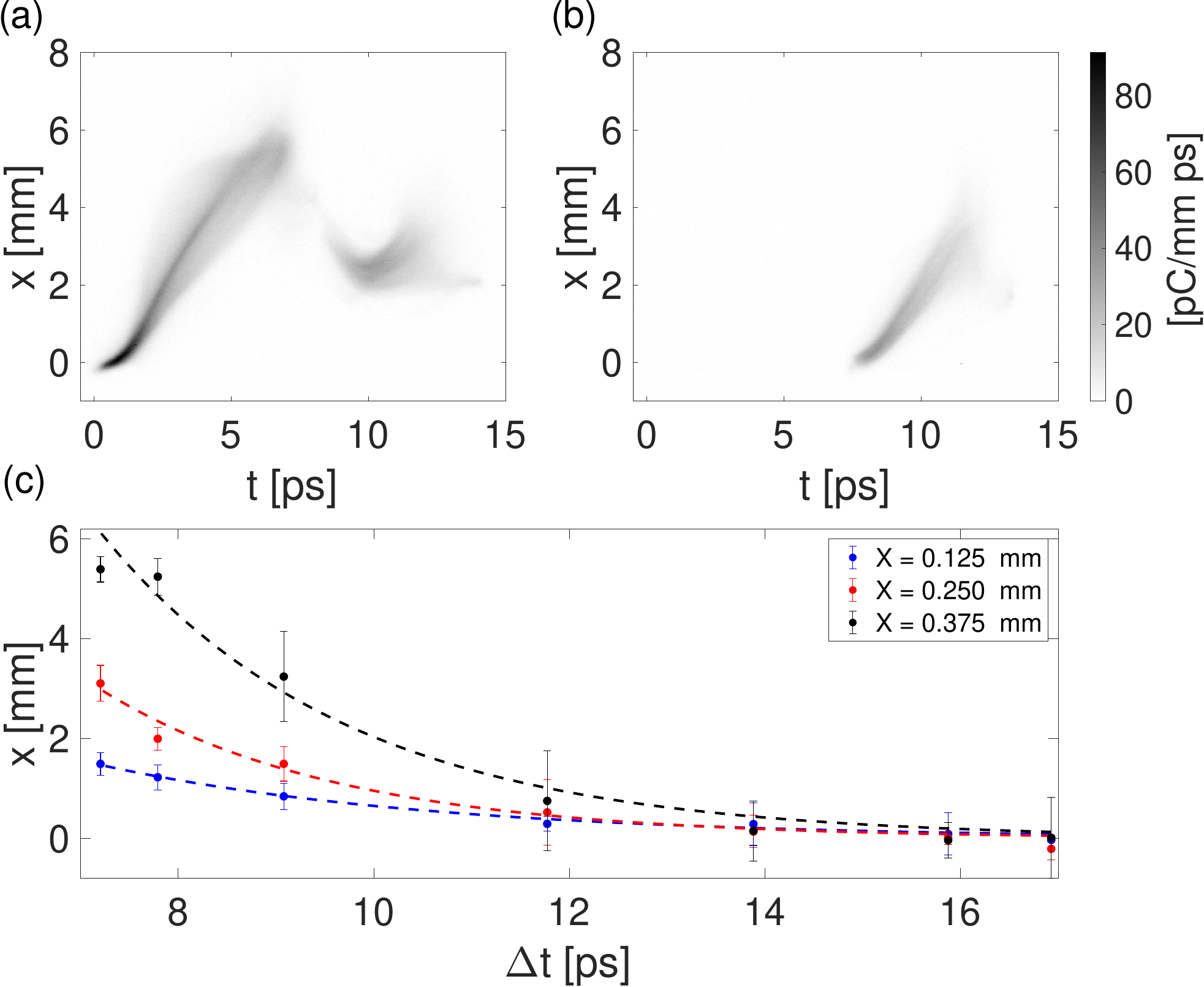}
\caption{
(a,b):  Single-event, time-resolved images of the 120\,pC trailing $e^-$ trailing witness bunch ($t>9.2\,$ps) with (a) and without (b) the 400\,pC drive $e^-$ bunch ($t>0\,$ps), respectively.
In both cases, horizontal offset $X=0.375\,$mm.
The bunches propagate from right to left.
(c): Transverse position of the front slice of the bunch as a function of the delay $\Delta t$ between the fronts of the two bunches for three misalignment positions. 
Dashed lines: results of exponential decay fit. 
}
\label{fig:3}
\end{figure}
\section{Experimental Results}

In Ref.~\cite{VERRA:2024}, we discussed the effect of the dielectric wakefields on the drive bunch only, and the conditions for which plasma screens the space-charge field of the bunch and therefore suppresses the dielectric wakefields.
When plasma is not present, or full space-charge field screening is not reached, the dielectric wakefields remain present behind the driver bunch for a finite amount of time.
The decay is more pronounced in this setup than in typical DWFA because of the absence of a metallic layer around the dielectric material~\cite{Ficcadenti:2018}.

\par Figure~\ref{fig:3}(a) shows a single-event, time-resolved image of the driver and trailing bunches (traveling from right to left) after propagation through the dielectric capillary with offset $X=0.375\,$mm.
The increasing steering effect of the dielectric wakefields along the drive bunch is clearly visible, as we already discussed in~\cite{VERRA:2024}.
In vacuum, the dielectric wakefields remain present behind the driver bunch ($t>0\,$ps) for a finite amount of time and they affect the trailing witness bunch ($t>9\,$ps).
\par The effect of the dielectric wakefields driven by the driver bunch is noticeable by comparing Fig.~\ref{fig:3}(a), where the delay between the front of the two bunches is $\Delta t = 9.1\,$ps, with (b), where the driver is not present. 
In the latter, the front of the trailing bunch propagates on its original trajectory ($x\sim0\,$mm), and the following part of the bunch is affected by the dielectric wakefields driven by the witness bunch itself.
The superposition of the two contributions (decreasing $W_{\perp}$ from the driver, increasing $W_{\perp}$ from the trailing bunch) results in the ``u" shape of the witness bunch in (a).

\par We measure the wakefields decay by measuring the displacement of the front of trailing witness bunch, whose delay with respect to the driver bunch is varied.
Figure~\ref{fig:3}(c) shows the transverse position of the front slice of the trailing bunch (where the amplitude of its own wakefields is negligible) as a function of the delay between the front of the two bunches, for three different misalignment values.
We note that the displacement is larger for larger misalignments, at any given $\Delta t$.
All cases show a decrease of the displacement with the increase of $\Delta t$ due to an exponential decay of $W_{\perp}$ (dashed lines show the result of fit with a function $\propto e^{-\Delta t/\tau}$).
This is in agreement with the expected behaviour of the surface charges generated by the passage of the driver bunch, and decaying afterwards~\cite{BANE:1984,COFFEY:2004}.
The results of these experiments show that the effect of $W_{\perp}$ is negligible (i.e., $x\sim 0$) for $\Delta t > 14\,$ps.
This can also be seen by comparing the
the centroid position of the back of the witness bunch ($t\sim 14\,$ps) in Fig.~\ref{fig:3}(a) with (b): the position is the same in the two cases because the wakefields driven by the driver bunch have already decayed. 

\par In a future PWFA based on dielectric capillaries, dielectric wakefields may be induced between consecutive driver-witness bunch couples, due to possible position jitters, when full space-charge field screening by the plasma does not occur. 
However, the results presented here show that the effect of these wakefield are negligible for $\Delta t >14\,$ps, 
that is much shorter than the recovery time measured in PWFA ($\mathcal{O}$(ns))~\cite{DARCY:2022,POMPILI:2024c}. 

\par In Ref.~\cite{VERRA:2024} we showed that the dielectric wakefields are suppressed when the distance between the bunch and the dielectric surface is much longer than the plasma skin depth. 
because the space-charge field of the $e^-$ bunch is screened by plasma.
The combination of these two results will be useful to design and optimize future PWFA based on dielectric capillaries. 
\section{Conclusions}

We measured the decay of the dielectric wakefields by measuring the deflection of a trailing witness bunch, and we found that the deflection effect becomes negligible for $\Delta t>14\,$ps, much shorter than the recovery time measured for plasma wakefields.
These results complement those discussed in a recent publication~\cite{VERRA:2024}, where we showed that the presence of plasma screens the space-charge field of the drive bunch and therefore suppresses the dielectric wakefields. 

\section*{Acknowledgments}
The work of L.V. has been supported by the European Union - Next Generation EU within the PNRR-EuAPS project.
This work has been partially by the European Commission under grant n. 101079773 (EuPRAXIA Preparatory Phase).
We thank G.~Grilli and T.~De Nardis for the development of the HV discharge pulser, F.~Anelli for the technical support and M.~Zottola for the experimental chamber installation.

 \bibliographystyle{elsarticle-num} 
 \bibliography{dielectric}





\end{document}